\documentclass[12pt,preprint]{aastex}
\usepackage{times}
\usepackage{color}
\usepackage{amsmath}
\usepackage{graphicx}
\usepackage{overpic}
\usepackage{enumerate}
\usepackage{subfigure}
\usepackage{multirow}
\usepackage{booktabs}
\usepackage{longtable}
\usepackage{threeparttable}
\usepackage{ulem}

\usepackage[breaklinks, colorlinks, urlcolor=red, filecolor=red, citecolor=blue, linkcolor=red, anchorcolor=red]{hyperref}

\newcommand{\es}{\,{\ensuremath{\rm erg\,{s}^{-1}}}}

\def\DP#1{{\ensuremath{10^{\rm #1}}}}
\def\TDP#1{{\ensuremath{\times 10^{\rm #1}}}}

\begin{document}

\title{A peculiar cyclotron line near 16 keV detected in the 2015 outburst of 4U\,0115+63?}
\author{Bai-Sheng Liu$^{1,2,3}$, Lian Tao$^{1,4}$, Shuang-Nan Zhang$^{1,4}$, Xiang-Dong Li$^{2,3}$, Ming-Yu Ge$^{1}$, Jin-Lu Qu$^{1,4}$, Li-Ming Song$^{1,4}$, Long Ji$^{5}$, Shu Zhang$^{1,4}$, Andrea Santangelo$^{5}$, Ling-Jun Wang$^{1}$}

\affil{$^{1}$Key Laboratory of Particle Astrophysics, Institute of High Energy Physics, Chinese Academy of Sciences, Beijing 100049, China; astrover2012@aliyun.com, taolian@ihep.ac.cn, zhangsn@ihep.ac.cn}
\affil{$^{2}$Department of Astronomy, Nanjing University, Nanjing 210046, China; lixd@nju.edu.cn}
\affil{$^{3}$Key Laboratory of Modern Astronomy and Astrophysics (Nanjing University), Ministry of Education, Nanjing 210046, China}
\affil{$^{4}$University of Chinese Academy of Sciences, Chinese Academy of Sciences, Beijing 100049, China}
\affil{$^{5}$IAAT, University of Tuebingen, Sand 1, Tuebingen D-72076, Germany}

\begin{abstract}

In 2015 October, the Be/X-ray binary 4U 0115+63 underwent a type II outburst, reaching an X-ray luminosity of $\sim$ {\DP{38}\es}. During the outburst, Nuclear Spectroscopic Telescope Array ({\it NuSTAR}) performed two Target of Opportunity observations. Using the broad-band spectra from {\it NuSTAR} (3$-$79 keV), we have detected multiple cyclotron lines of the source, i.e., $\sim$ 12, 16, 22 and 33/35 keV. Obviously, the 16 keV line is not a harmonic component of the 12 keV line. As described by the phase-dependent equivalent widths of these cyclotron lines, the 16 keV and 12 keV lines are two different fundamental lines. In our work, we apply the two-poles cyclotron line model to the observation, i.e., the two line sets are formed at the same altitude ($\sim$ 0.2 km over the NS surface) of different magnetic poles, with $\sim$ {1.1\TDP{12}} and {1.4\TDP{12}} G in two poles, respectively.

\end{abstract}

\keywords{UAT: Accretion (14); Neutron stars (1108); High mass X-ray binary stars (733); Binary pulsars (153)}

\section{Introduction}
\label{sec01}

In binary systems, by accreting mass from their companion stars, highly magnetized ($B\gtrsim$ {\DP{12}} G), rotating neutron stars (NSs) are observed as X-ray pulsars. The interaction between the accretion flow and the magnetic field is one of the key issues in X-ray astronomy. On one hand, the accretion flow tends to be channeled onto the polar cap of the NS along the field lines, and a hotspot or thermal mound is formed, which contributes to a large fraction of X-ray emission \citep[][]{BeckerW07}. On the other hand, due to the strong field the energies of electrons are quantized according to the Landau level \citep[][]{Meszaros92}
\begin{eqnarray}
  E_{n}=n\frac{e{B}\hbar}{m_{\rm e}c} \approx 11.6{\,\rm keV}\cdot{n}B_{\rm 12},\label{eqn01}
\end{eqnarray}
where $m_{\rm e}$ is the mass of an electron, $n=$1, 2, 3 $\ldots$ and $B_{\rm 12}=B/10^{12} {\,\rm G}$. The cyclotron resonant scattering feature (CRSF, or commonly referred to as `cyclotron line') of the outgoing photons with these quantized electrons happens in the line-formation region, which naturally explains the absorption features observed in the energy spectra of $\sim$ forty X-ray pulsars \citep[e.g., ][]{RevnivtsevM15, StaubertTK19, TruemperPR78, WalterLB15, WheatonDP79}. If the gravitational redshift is considered, the observed energy of the cyclotron line should be $E_{\rm cyc}=E_{n}/(1+z)$, where $z$ is the gravitational redshift around the NS. At present, it is the only way to directly measure the magnetic field of NSs. However, the exact location of the line-formation region is still under discussion, and the cyclotron line is occasionally absent in some pulsars. These issues inhibit us from revealing the evolution of $E_{\rm cyc}$ with the X-ray luminosity \citep[e.g., ][]{DoroshenkoTME17, StaubertKE17, StaubertTK19, TsygankovLCS07}.

Furthermore, besides the fundamental line, multiple harmonics were exhibited in the spectra of several X-ray pulsars, e.g., 4U 0115+63 \citep{SantangeloSG99}, V 0332+53 \citep{TsygankovLCS06}, and GRO J2058+42 \citep{MolkovLT19}. From these harmonics, we can infer the fundamental line by calculating the minimal energy gap between them, which is of vital importance to the measurements of the magnetic fields and further studies on X-ray pulsars. In our work, we mainly study the multiple absorption features of 4U\,0115+63.

4U\,0115+63, locating at $\sim 7$ kpc \citep{NegueruelaO01,BailerJRFE18}, is a Be/X-ray binary \citep[][]{JohnsKCM78}, and its orbital parameters have been measured \citep[orbital period $P_{\rm orb}\sim$ 24.3 d, eccentricity $e\sim$ 0.34,][]{RappaportCCL78}. It consists of a $\sim$ 3.61 s accreting pulsar \citep[][]{CominskyCLM78} and a Be star. Like other transient Be/X-ray binaries, 4U\,0115+635 had undergone several outbursts, during which the fundamental line and several harmonics were detected \citep[e.g., ][]{BoldinTL13, FerrignoBSM09, LiWZ12, SantangeloSG99, TsygankovLCS07}. Especially, for the first time a peculiar cyclotron line near 15 keV was probed simultaneously with those near 11, 20 and 33 keV, and was considered as the other fundamental line \citep[][]{IyerMD15}. If two different fundamental lines are detected simultaneously, it is unclear where they are produced, and from which one the magnetic field can be measured. However, the robustness of the peculiar absorption feature needs further confirmation. On one hand, the signal to noise (S/N) of the observation data, obtained from the joint observations of {\it Suzaku} and {\it RXTE}, is somewhat low, since different systematic errors introduce extra uncertainties. On the other hand, the 11 keV line is detected at a low level of significance (3.5$\sigma$), which makes the coexistence of the two line sets doubtful. That is, observations of both line sets with a higher S/N with the same telescope are needed.

In this work, by analyzing two pointing observations of 4U 0115+63 during the 2015 outburst, obtained by the Nuclear Spectroscopic Telescope Array ({\it NuSTAR}), we try to study the complicated cyclotron lines. {\it NuSTAR} performs X-ray observations in 3$-$79 keV, displaying energy resolution of 400 eV (full width at half maximum) at 10 keV and 900 eV at 68 keV. Its inherently low background makes the telescope detect hard X-ray sources with an improvement in sensitivity by 100 times as high as other instruments \citep[][]{HarrisonCE13}. Its time resolution (2 $\mu$s) and dead time (2.5 ms) further allow us to obtain observation data with high quality and to analyze pulse phase-resolved energy spectra conveniently. Thus these two {\it NuSTAR} observations are helpful to confirm the robustness of the peculiar CRSF and reveal its nature. From phase-dependent equivalent widths of the measured CRSFs and the physical model of the accretion column \citep{BeckerW07, BeckerKSE12}, we can infer the formation processes of these lines. In Section \ref{sec02}, we describe the observation and data reduction. In Section \ref{sec03}, we present details of the spectral fitting, and test the robustness of the peculiar cyclotron line. In Section \ref{sec04}, we try to explain the nature of these confusing absorption lines, and summarize our work.

\section{Observation and Data Reduction}
\label{sec02}

\begin{figure*}
\begin{center}
  \includegraphics[width=0.5\textwidth, angle=0]{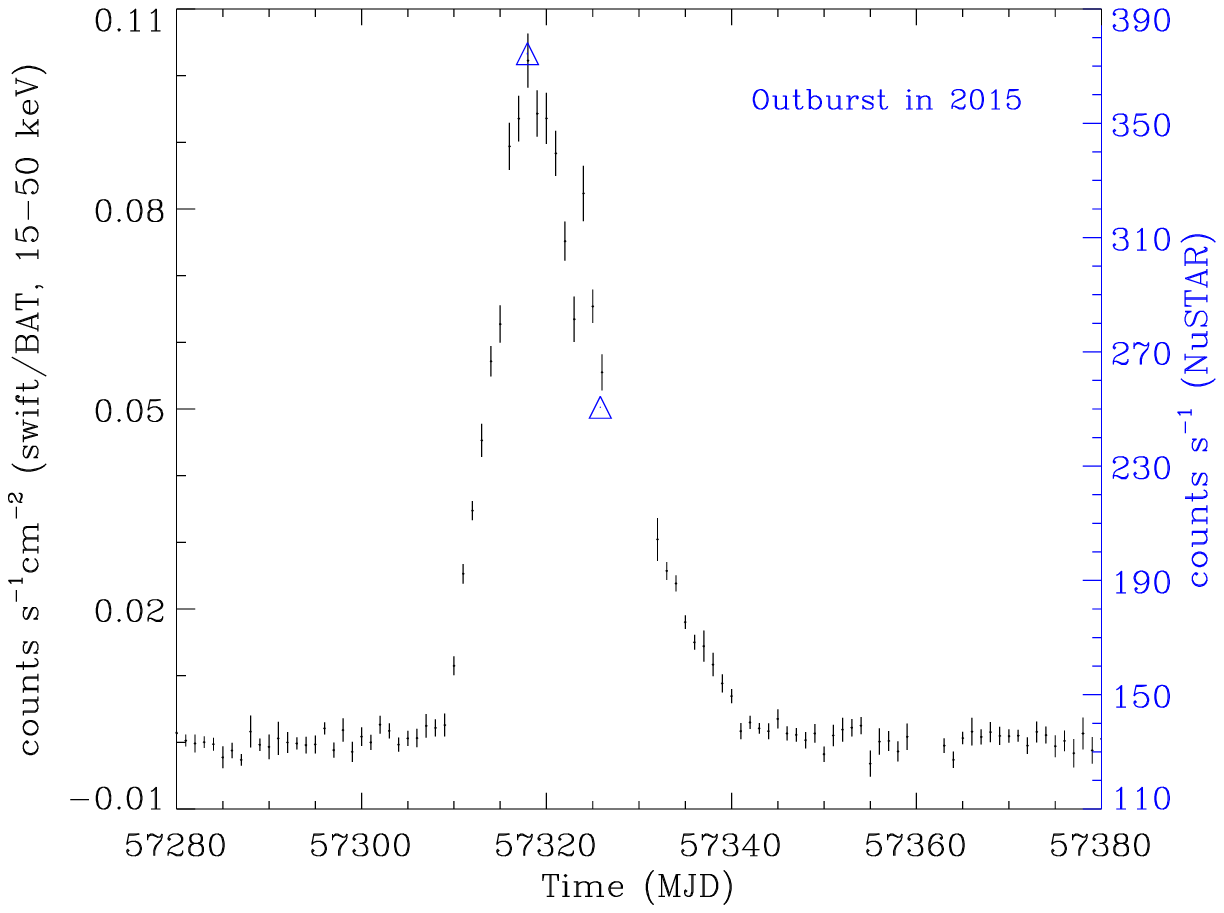}
  \caption{The 2015 outburst of 4U 0115+63 monitored by {\it Swift}/BAT (black points). Two {\it NuSTAR} observations used in this paper are marked in blue triangles.}
  \label{fig01}
\end{center}
\end{figure*}

During the 2015 outburst of 4U 0115+63, {\it NuSTAR} performed a Target of Opportunity (ToO) observation of 4U\,0115+63 on 2015 October 22 and 30 (ObsIDs 90102016002 and 90102016004, hereafter abbreviated as ObsIDs 002 and 004), respectively. The data is obtained with the focal plane module telescopes (FPMA and FPMB). The net exposure times of the two observations are $\sim$ 8.6 and 14.6 ks, respectively. Figure \ref{fig01} displays the details of these ToO observations during the 2015 outburst (described by {\it Swift}/BAT), i.e., these {\it NuSTAR} observations are performed in the peak and decay of the outburst, respectively.

\clearpage

Following the analysis guide of {\it NuSTAR} data, we first employ the {\tt nupipeline} task (v 0.4.6) of the software {\tt NUSTARDAS} (v 1.8.0, packaged in {\tt HEASOFT} v 6.23) to filter and calibrate the event data, using {\it NuSTAR} Calibration data base (CALDB; released on Oct. 22 2018)\footnote{Here we execute the tasks of {\tt nupipeline} and {\tt nuproducts} two times. In the repeated procedure, the keyword {\tt statusexpr} of ``STATUS==b0000xxx00xxxx000" is considered since in the preliminary processing the incident rate of the light curves with binsize=1 s exceeds 100 counts per second (see \url{https://heasarc.gsfc.nasa.gov/docs/nustar/nustar_faq.html}).}. Secondly, we utilize the {\tt nuproducts} task to extract source (and background) light curves and spectra from the cleaned FPMA and FPMB data, and response files. We obtain the source products using a circular region of 160 arcsec around the source position, and the background products using a circular region of 130 arcsec around a position away from the source region. Thirdly, we group the FPMA and FPMB energy spectra via the {\tt grppha} task with $\geq$ 50 counts per channel bin.

\section{Spectral Analysis}
\label{sec03}

\subsection{Spectral Fitting}
\label{sec03.1}

To study the cyclotron absorption of the Be/X-ray binary 4U 0115+63 in the giant outburst, we analyze the phase-averaged spectra in 3$-$79 keV. For ObsID 002 or 004, the FPMA and FPMB energy spectra are fitted together, and the cross normalization factors are adopted, i.e., the factor for FPMA is fixed at 1.0, and that for FPMB is free \citep[with uncertainty of 1$-$2\%, in good agreement with][]{MadsenHM15}. Other parameters for both spectra are tied, respectively.

The broad-band spectra are tentatively fitted with the commonly discussed models. In order to find the interstellar absorption, we adopt the {\tt TBabs} model in {\tt XSPEC} (v 12.10.0) by setting abundances and cross-sections in accordance with those of \citet{WilmsAM00} and \citet{VernerY95}, respectively. Since the calibrated spectra below 3 keV are not available here, we fix the absorption column ($N_{\rm H}$) at {1.2\TDP{22}} cm$^{-2}$ in all fits unless specified \citep[$\sim (0.5-2)$ {\TDP{22}} cm$^{-2}$, see][]{FerrignoBSM09, IyerMD15, TsygankovLD16}. The almost simultaneous {\it swift}/XRT observations (on Oct. 21, 23 and 30) are not considered in our work. It is shown that $N_{\rm H}$ can not be well determined from the jointly spectral fitting. In that there are some inconsistent calibration, different scattering halos (producing different values of $N_{\rm H}$) between two telescopes\footnote{See \url{http://iachec.scripts.mit.edu/meetings/2019/presentations/WGI_Madsen.pdf}}. To fit an absorption feature, we can use a multiplicative gaussian \citep[In {\tt XSPEC} we define a multiplicative model {\tt mgabs} using the function \ensuremath{1-\tau\exp[-\frac{(E-E_{\rm cyc})^2}{2\sigma^2}]}. See][]{MiharaMOS90, FerrignoBSM09, DoroshenkoTME17, StaubertTK19}, a fake-Lorentzian ({\tt cyclabs}) or an exponential gaussian model ({\tt gabs}) \citep[e.g., ][]{NakajimaMM10, MullerFK13, StaubertTK19}. Here we only adopt model {\tt mgabs} since the other two models may cause some shift in the measured line center or some residuals around the fitted line \citep[see discussions in][]{DoroshenkoTME17}.

First, we fit the spectra of ObsID 002 using a {\tt compTT} model. As shown in the left panel `a' of Fig. \ref{fig02}, a significant soft excess is always detected, even when different values of $N_{\rm H}$ ($\sim (0.5-2)${\TDP{22}} cm$^{-2}$) are considered. Adding a low temperature {\tt bb} component significantly improves the spectral fit with $\Delta \chi^{2} \gtrsim 11000$. An emission line near 6.5\,keV, three narrow absorption lines ($\sim$ 23, 35 and 48\,keV) and a broad absorption feature ($\sim 10-20$\,keV) become apparent in the residuals (see the left panel `b'). We then add a {\tt gaussian} and four {\tt mgabs} components to fit these features, which leads to a substantial decrease in $\chi^{2}$ (see the left panel `c'). Upon closer inspection, we find two small dips in the residuals near 12\,keV and 16\,keV. The line features are likely related to the poor modelling of the absorption feature in $10-20$\,keV. We suspect that there are two narrow absorption lines rather than a broad one. Therefore, we test the fit by adding another line, and finally obtain a very good fit (see the left panel `d' and right panel of Fig. \ref{fig02}) with $\chi^2$/dof $=1828/1731$ (degree of freedom) and $\Delta\chi^2 \sim$ 100. In summary, we have detected four harmonics ($\sim$ 12, 23, 35 and 48 keV) and a peculiar one \citep[$\sim$ 16 keV, not a harmonic component of the 12 keV line. See also][]{RoyAIB19} in ObsID 002, even when we adopt different values of $N_{\rm H}$ ($\sim(0.5-2)$ {\TDP{22}} cm$^{-2}$). Especially, our further analysis precludes the residual near 12 keV related to the {\it NuSTAR} calibration issue \citep[for more details see][]{DoroshenkoPDS20}. If each CRSF is fitted by model {\tt cyclabs} or {\tt gabs}, the absorption feature near 16 keV can still be identified.

Then we repeat the above procedure using some power-law based models \citep[phenomenological ones, see][]{MullerFK13, StaubertTK19}, e.g., the simple cutoff power-law ({\tt cutoffpl}), a power-law with a high energy cutoff (HEcut), that with a `Fermi-Dirac' cutoff (FDcut), or a sum of a negative and positive power-law multiplied by an exponential cutoff (NPEX). We still detect the abnormal CRSF near 16 keV in the fit using any of these power-law based models (see Fig. \ref{fig03}), even when model {\tt cyclabs} or {\tt gabs} is applied. If the residual near 16 keV is additionally fitted as a CRSF, the fit is obviously improved as compared to that in Fig. \ref{fig03}, e.g., $\chi^2$ of model {\tt cutoffpl}, NPEX, HEcut and FDcut based fittings are reduced by $\sim$ 109, 131, 53 and 357, respectively.

Similarly, in ObsID 004, aside from the harmonic CRSFs ($\sim$ 12, 22 and 33 keV), a peculiar one near 16 keV is again detected (see Tbl. \ref{tab01}), when each model used in ObsID 002 is considered.

However, the absorption feature at $\sim$ 16 keV might be related to the so-called ``10 keV feature'', described by a wide-{\tt gaussian} profile \citep[e.g., ][]{Coburn01, FerrignoBSM09, MullerFK13, FarinelliFBB16, StaubertTK19, neKuhnelKFE20} or a {\tt compTT} component \citep[e.g.,][but not applicable to 4U 0115+63, since in the tentative fitting the ``10 keV feature'' still exists]{TsygankovDM19, TsygankovES19}. That is, if each of the above fits contains an additive wide-{\tt gaussian} component (see Tbl. \ref{tab01}), the peculiar absorption line at $\sim$ 16 keV would not be detected (see also Fig. \ref{fig04}). Thus, the detected absorption lines are all harmonic, e.g., $\sim$ 12, 23 and 33/35 keV. Even though in the fit including a wide-{\tt gaussian} component the statistical $\chi^2$/dof decreases a little bit accordingly, the wide-{\tt gaussian} profile is associated with unknown physics, besides reflecting the possible distribution of the thermal atoms \citep[e.g., ][]{StaubertTK19}. In the following, we treat the residual near 16 keV as a CRSF, unless future observations can obtain some astrophysical evidence relating to the wide-{\tt gaussian} model.

\subsection{Robustness of the 16 keV line}
\label{sec03.2}

\begin{figure*}
  \centering
  \begin{minipage}{7cm}
	\centering
	\includegraphics[scale=0.6, angle=0]{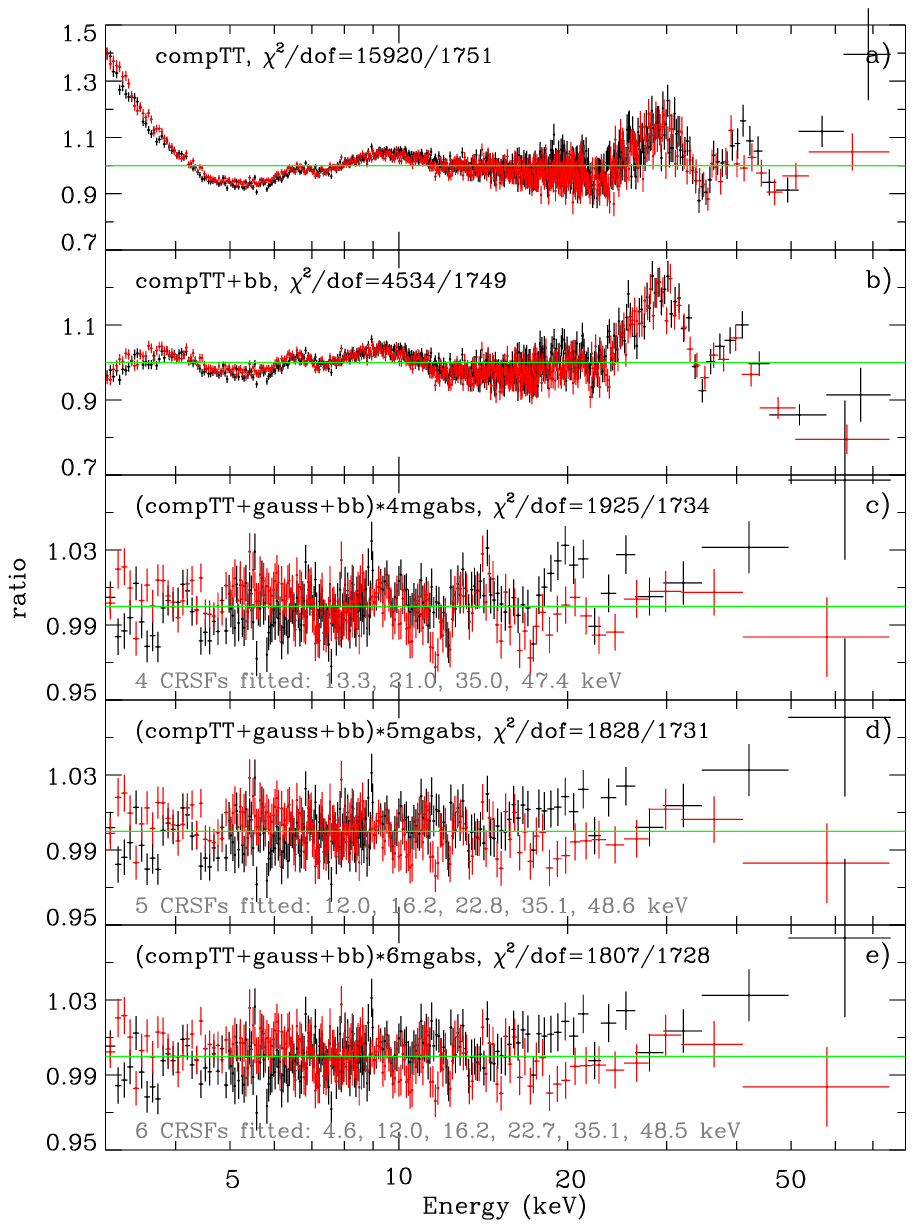}
  \end{minipage}
  \begin{minipage}{7cm}
	\centering
  \begin{overpic}[scale=0.4, angle=-90]{f02b.ps}
  \end{overpic}
  \end{minipage}
  \caption{Left panels: spectral residuals with respect to different models for ObsID 90102016002. Panel `d' is the residuals for the best-fitting model, and panel `e' is that of the tentative fitting by adding a CRSF near 5 keV. A {\tt compTT} model is used to fit the continuum. Right panels (from top to bottom): energy spectra and model components; spectral residuals with one line from the 5 CRSFs (12.0, 16.2, 22.8, 35.1 and 48.6 keV) removed, respectively. Except as specifically described, the following figures show the results from ObsID 90102016002.}
  \label{fig02}
\end{figure*}

\begin{figure*}
\begin{center}
  \includegraphics[width=0.45\textwidth, angle=-0]{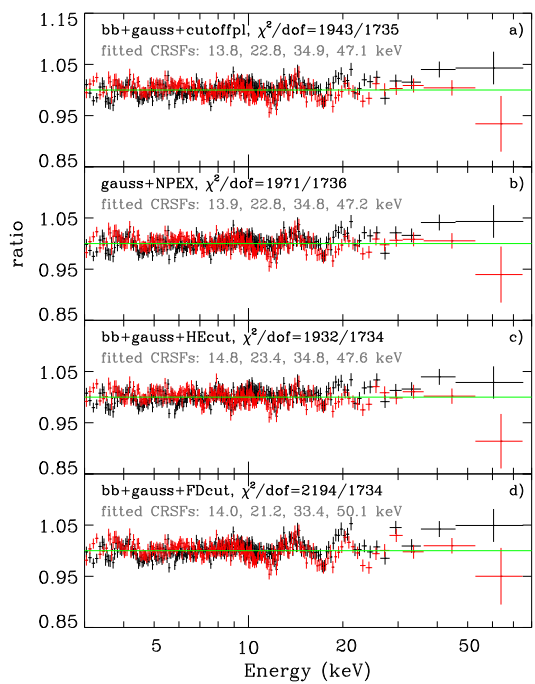}
  \caption{Spectral residuals of different continuum models ({\tt cutoffpl}, {\tt NPEX}, {\tt HEcut} or {\tt FDcut}). In order to confirm the presence of the residual near 16 keV, we test different continuum models. In all cases, the line feature around 16 keV is always apparent.}
  \label{fig03}
\end{center}
\end{figure*}

In the above analyses, we have described the details about the detection of multiple cyclotron lines in 4U 0115+63. Especially, we have detected an anomalous CRSF ($\sim$ 16 keV) in both {\it NuSTAR} observations. Although the similar multiple CRSFs (especially that near 15 keV) in the 2011 outburst have been reported by \citet{IyerMD15}, due to higher performance of {\it NuSTAR}, the CRSFs in our work are measured with a higher S/N. Obviously, the 16 keV line can not be treated as a harmonic component (the fundamental) of the 12 keV (23 keV) line. It seems that the cyclotron line near 16 keV is a different fundamental line, unless an absorption line near 5 keV exists\footnote{Note that some weak absorption-like residual near 5 keV appears in the spectrum of ObsID 002, regardless of the fitting model (see Fig. \ref{fig02}$-$\ref{fig04}). According to our further spectral analysis, we cannot treat the residual as an absorption feature, since the tentative fitting displays no significant improvement ($\Delta\chi^2 \sim 20$, see left panels `d' and `e' of Fig. \ref{fig02}), and the width of the 4.6 keV line is $\lesssim$ 0.1 keV. In addition, the small residual may be related to the uncertainty of the calibration \citep[][]{MadsenHM15, MadsenFE17}.}, or the 16 keV line is a minor product of the 12 keV line. Therefore, the nature of these two sets of cyclotron lines should be studied, e.g., whether they are formed in the same region or not. Then the cyclotron line near 35 keV (48.5 keV) may also be the first (second) harmonic of the 16 keV line, besides being the second (third) harmonic of the 12 keV line. In Section \ref{sec04.1}, by analyzing the phase-dependent equivalent widths of these cyclotron lines and physical model of the accretion column, we try to answer these questions.

Before modelling the 12 and 16 keV lines, we should check their robustness of detection. By fitting only four (three) CRSFs in ObsID 002 (004), we can roughly estimate the significance from observation data. Our fit tends to fit CRSFs near 12 and 16 keV rather than the harmonics (i.e., $\sim$ 12 and 23 keV), since the former scheme produces a smaller value of $\chi^2$/dof or a group of satisfactory parameters. For example, using the {\tt compTT} dependent model in ObsID 002, we obtain $\chi^2$/dof of 1851.3/1734 and 1924.78/1734 for the scheme fitting the lines at 11.9, 16.2, 23.2 and 34.8 keV and that fitting the lines at 13.3, 20.8, 35.1 and 47.5 keV, respectively. That is, the absorption features near 12 and 16 keV should not be ignored (see also the right panel of Fig. \ref{fig02}).

In order to further confirm the robustness of the CRSF near 16 keV, we perform the following simulations and fitting-statistics. Briefly, the 16 keV line is not involved to produce the simulated spectra, and due to the statistical fluctuations its possible detection from the simulated spectra is analyzed. The details are described as follows. (i) Basing on the physical model {\tt compTT+gauss+bb} (modified by {\tt TBabs} and four {\tt mgabs} components), we simulate {\DP{5}} spectra using script {\tt fakeit} in {\tt XSPEC} package. Besides the response matrix files and auxiliary response files (for FPMA and FPMB) of ObsID 002, the best parameters for the spectrum in Tbl. \ref{tab01} (except those of the 16 keV line) are applied. Additionally, the exposure durations of ObsID 002 (8.58 and 8.88 ks for FPMA and FPMB, respectively) and statistical (Poisson) noise are considered to produce simulated source/background spectra. (ii) Applying model {\tt compTT+gauss+bb} absorbed by {\tt TBabs} and four {\tt mgabs} components, we fit the simulated spectra and obtain their $\chi^2$. By including another {\tt mgabs} component with an initial trial $E_{\rm cyc}=$16.20 keV (varying between 15 and 17 keV) and a fixed line width of 3.17 keV, we again fit the spectra, and calculate $\Delta\chi^2$ as compared to the former fit. Then the number distribution of these $\Delta\chi^2$ is shown in panel `a' of Fig. \ref{fig05}. As illustrated in the figure, none of these simulated spectra displays a $\Delta\chi^2$ close to that of the observation ($\sim -$ 103.8). That is, if four harmonic lines ($\sim$ 12, 23, 35 and 48 keV) are detected in the observed data, the probability of nonexistence of the 16 keV line is lower than {\DP{-5}}.

Then we repeat the above simulation to verify the robustness of the 12 keV line, and obtain a similar $\Delta\chi^2$ distribution in panel `b' of Fig. \ref{fig05}. In the energy band 11$-$13 keV of the simulated spectra, we can not detect an assumed 12 keV line with $\Delta\chi^2$ close to that of the observation ($\sim -$ 173.0). Therefore, the simultaneous detection probability of the 12 keV line with other four cyclotron lines ($\sim$ 16, 23, 35 and 48 keV) is greater than 99.999\% in the observed data.

\section{Discussion and conclusion}
\label{sec04}

In 4U 0115+63, two cyclotron lines near 12 keV and 20 keV were first detected by {\it HEAO} in the 1978 outburst \citep{WheatonDP79, WhiteSH83} and again in the 1990 outburst \citep[by {\it GINGA},][]{MiharaMN98}, and a single line around 17 keV appeared in the 1991 outburst. In different epochs of the 1999 outburst the second ($\sim$ 33 keV), third ($\sim$ 49 keV) and fourth ($\sim$ 57 keV) harmonics were obtained by {\it Rossi X-Ray Timing Explorer} \citep{HeindlCGE99}, {\it BeppoSAX} \citep{SantangeloSG99} and {\it BeppoSAX} \citep{FerrignoBSM09}, respectively. These absorption lines were further confirmed in more outbursts, observed by other detectors \citep[e.g.,][]{BoldinTL13, LiWZ12}. In the 2011 outburst, among detected CRSFs ($\sim$ 11, 15, 20 and 33 keV), the 15 keV line is not a harmonic of the 11 keV line \citep{IyerMD15}, and the similar situation happened again in the 2015 outburst \citep[see this work and][]{RoyAIB19}.

In order to reveal the nature of these complicated cyclotron lines observed in the 2015 outburst, we study the pulse phase-resolved spectra. Basing on {\it NuSTAR} observations, we obtain the phase-resolved spectra in 10 phase bins with equal width, and fit these spectra using the {\tt bb+gauss+compTT} based model. During the fitting, we assume the detections of all CRSFs in each bin by fixing their widths to those of the phase-averaged spectra, respectively, beside fixing $\sigma_{\rm Fe}$ accordingly (see Tbl. \ref{tab01}). In a few cases, we also fix the line energy if its upper and lower limits can not be constrained simultaneously (see the data point with a downward arrow in Fig. \ref{fig06}). Then we calculate the phase-dependent equivalent widths (EWs) of the CRSFs (see Fig. \ref{fig06}. Since the line near 48 keV is not detected in ObsID 004, we only study other CRSFs) using the following equation (based on model {\tt mgabs}),
\begin{eqnarray}
  {\text{EW (keV)}}= \int^{E_{\rm 2}}_{E_{\rm 1}} \tau\exp[-\frac{(E-E_{\rm cyc})^2}{2\sigma^2}]{\rm d}E,\label{eqn02}
\end{eqnarray}
where $\tau$ and $\sigma$ are the line depth and width, respectively.

For further discussions, we pick up two energy-dependent pulse profiles (8$-$14 and 14$-$19 keV), each of which is affected by the cyclotron absorption (see the profiles without the cyclotron absorption and those with the absorption in panels `N' and `a' of Fig. \ref{fig06}, respectively). In each pulse phase, using the script {\tt cflux} in {\tt XSpec} we first calculate the flux ($F_{\rm 1}$, e.g., in 8$-$14 keV) without the cyclotron absorption (e.g., at 12 keV) and that ($F_{\rm 2}$) with the absorption, respectively. Then, in each phase multiplying the count rate by the factor $F_{\rm 1}/F_{\rm 2}$, we can obtain the pulse profile without the cyclotron absorption in 8$-$14 keV. Similarly, we can work out the profile without absorption in 14$-$19 keV. We also plot the phase-dependent $E_{\rm cyc}$ of the 12 keV and 16 keV lines in panel `d'.

As depicted in Fig. \ref{fig06}, $E_{\rm cyc}$ (phase varied by up to 30\%) and `EWs' all show significant pulse-phase dependence. Obviously, every CRSF is detected at most of the pulse phases. The phase-dependent $E_{\rm cyc}$ and the corresponding pulse profile reach their peaks almost at the same phase, which is similar to Her X-1 \citep[][]{StaubertKE14}. The details about the phase-dependent EWs are summarized as follows.
\begin{enumerate}[1)]
  \item{The line-formation region of the 12 keV (16 keV) line is primarily observed at $\phi \sim$ 0.6$-$0.9 (0.2$-$0.5), at which the pulse profile for the 8$-$14 keV (14$-$19 keV) band has a hump, and displays significant cyclotron absorption (see panels `N' and `a' of Fig. \ref{fig06}). At $\phi \sim$ 0.9$-$1.1 the 16 keV line has another peak EW (e.g., in ObsID 004), where no obvious pulse is detected. Additionally, EW of the 16 keV line is larger than that of the 12 keV line.}
  \item{As for the first harmonic of the 12 keV line, the 22 keV line reaches its peak EW at $\phi \sim$ 0.2$-$0.5 and 0.6$-$0.9. Especially, at $\phi \sim$ 0.2$-$0.5 the significant absorption near 22 keV conflicts with the weakness of the 12 keV line (see ObsID 004), different from the situation at $\phi \sim$ 0.6$-$0.9.}
  \item{As for the second (first) harmonic of the 12 keV (16 keV) line, the line near 33/35 keV displays very strong absorption at $\phi \sim$ 0.1$-$0.2, 0.5$-$0.6 and 0.8$-$0.9, and very weak one at $\phi \sim$ 0.2$-$0.5. At the phase where the strong absorption near 33/35 keV appears, no component of the 12 keV or 16 keV line set arrives at its peak EW.}
\end{enumerate}

Therefore, we believe that two sets of cyclotron lines (fundamental lines of $\sim$ 12 keV and 16 keV, respectively) coexist in the 2015 outburst. First, the formation of the 12 keV and 16 keV lines are not affected by each other. Until now no evidence indicates that an absorption feature near 5 keV is detected (i.e., the two lines are not harmonic), or that the 16 keV line is the by-product of the 12 keV line (see discussions in Sec. \ref{sec03.2}). Secondly, the 12 keV and 16 keV lines should be formed in different regions, since their EWs show different pulse phase dependence, respectively (see Fig. \ref{fig06}).

We also estimate the systematic uncertainty related to the fixed cyclotron line width. During the spectral fitting, we first let the line width range from 0.9 to 1.1 times of the width obtained from the phase-averaged spectrum. After obtaining the best value of the width, we fix the width and calculate the errors of other CRSF parameters. Then it is shown that EWs in most phases would be varied by $\lesssim$ 30\%. Even so, the above main conclusions are hardly affected.

In the following, we further discuss the formation of these different line sets using the physical model of the accretion column.

\subsection{About the absorption feature near 16 keV}
\label{sec04.1}

In the previous work analyzing the 2011 outburst \citep{IyerMD15}, the complicated cyclotron lines, classified into two different line sets, are believed to be formed in different regions, which is in consistence with our analysis. They proposed two possibilities to explain the detected lines.

The first case depends on the dipolar structure of the NS magnetic field, where the 11 and 15 keV line sets are formed in the pencil beam (the top of the shock) and fan beam (the base of the hot mound), respectively. The model can describe the formation of the two possible sets of cyclotron lines in GX 301$-$2 \citep[][]{FurstFME18}. However, the model is in disagreement with the bright outburst \citep[see Eq. (32) in][]{BeckerKSE12}\footnote{For ObsIDs 002 and 004, the 0.1$-$100 keV luminosities are, respectively, {1.08\TDP{38}\es}$(d/\text{7 kpc})^2$ and {7.28\TDP{37}\es}$(d/\text{7 kpc})^2$. In the 2011 outburst, the 3$-$50 keV luminosities $\gtrsim$ {2\TDP{37}\es}.}. First, the pencil beam can not be formed at the top of the accretion column where upward photons are trapped by the advection of the accretion flow. Moreover, the column top \citep[$\sim$ 10 km, see Eq. (16) in][]{BeckerKSE12} is higher than the location ($\sim$ 1.0 km, assuming $B \propto R^{-3}$ and on NS surface $E_{\rm cyc}\simeq$ 16 keV, where $R$ is the distance to the NS center) at which the 12 keV line set is formed. Therefore in the bright state it is not reasonable to have two different line-formation regions on the same pole.

In the second case, \citet{IyerMD15} supposed that the two sets of CRSFs are formed on two poles of the NS with nondipolar magnetic fields, respectively, i.e., two-poles CRSF model (TPCM). The viability of TPCM depends on two key points. First, the magnetic fields of the two poles should be different, i.e., a nondipolar field. By decomposing the energy-dependent pulse profiles of 4U 0115+63 at different levels of luminosities, \citet{SasakiMKF12} derived that the magnetic axes of the two poles are misaligned (offset by $\sim$ 60$^{\rm o}$). In the distorted configuration, it is reasonable to assume that the local magnetic fields of the two poles are different. Secondly, two groups of cyclotron absorption should happen in the two poles, respectively. Given that each energy-dependent pulse profile ($\lesssim$ 50 keV) has double peaks (a main and minor one), \citet{IyerMD15} analyzed the energy-dependent phase lags of each peak by defining a reference pulse profile \citep[see also][]{FerrignoFBB11}. The significant negative phase-shifts of the main (minor) peak are detected at energies of $\sim$ 11, 23 and 39 keV (16 and 30 keV), respectively, indicating the energies at which the corresponding cyclotron absorption happens. Provided that each peak in the pulse profile of 4U 0115+63 mainly corresponds to the emission from one single pole, they concluded that the two sets of CRSFs are formed in the two poles, respectively. However, their method in testing the second point of TPCM is inconsistent with some observation, e.g., some energy-dependent pulse profile displays more than two peaks (see panel `002-a' of Fig. \ref{fig06}), from which it is ambiguous to infer the phases of the two poles in the profile. In addition, the two line sets might be produced in the fan and pencil beams, respectively, if these radiation regions contribute to different humps in the pulse profile accordingly \citep[e.g.,][]{SasakiMKF12, IwakiriPE19}. Especially, in the bright state ($\gtrsim$ {\DP{37}\es}), neither of the two peaks in the profile corresponds to the emission of a single pole \citep{KrausBSE96}.

Thus in the following we apply some other way to test the second point in TPCM. In our work, not only does the measurement with higher S/N identify the detection of the peculiar cyclotron line near 16 keV, but also we can obtain phase-resolved spectra due to the high quality of {\it NuSTAR} data, which supplies more details about these CRSFs. That is, basing on the phase-dependent `EWs' of these CRSFs (i.e., the strength of the cyclotron absorption in different phase, see Fig. \ref{fig06}), we can constrain better their line-formation regions. Then we check whether these two regions can be recognized as the two magnetic poles, respectively. Our analyses concentrate on the dependence of the `EW' on the pulse profiles for the low energy band ($8-14$ keV and $14-19$ keV), irrespective of the appearance of more than two peaks in the profiles, or the contribution of different radiation regions to the pulses.

As described by the following discussions, the observations in Fig. \ref{fig06} are consistent with the second point in TPCM. As depicted in panels `N' and `a' of Fig. \ref{fig06}, the continuum radiation of 8$-$14 keV (14$-$19 keV) undergoes a significant cyclotron absorption at $\phi \sim$ 0.6$-$0.9 (0.2$-$0.5), which denotes the line-formation region of the 12 keV (16 keV, see panel `b') line. In TPCM, these regions can be identified as two different magnetic poles, respectively (see the shade region in panel `a'). Moreover, at the phase where the 12 keV (16 keV) line is formed, we also witness the slightly weaker absorption near 16 keV (12 keV). According to \citet[][]{KrausBSE96}, in high-luminosity state ($L_{\rm X} \gtrsim$ {\DP{37}\es}), the emissions from two magnetic poles can both contribute to the formation of each pulse. Therefore in the same pulse both two sets of cyclotron lines appear (see panels `a' and `b').

We can further determine which line set is formed on pole 1 and the other is on pole 2. It is the scattering cross-section of electrons that affects the interaction of electrons with photons and the propagation of photons. In strong magnetic field ($\gtrsim$ {\DP{12}} G), the cross-section for low-energy photons ($E <E_{\rm cyc}$) depends on the field strength, propagation angle with respect to the field and photon energy \citep[][]{AronsKL87, BeckerW07, BeckerKSE12}. The cross-sections parallel ($\sigma_{\rm \|}$) and perpendicular ($\sigma_{\rm \bot}$) to the magnetic field can be described, respectively, by \citep[see discussions in][]{BeckerW07}
\begin{eqnarray}
  &&\sigma_{\rm \|} \approx (\frac{E}{E_{\rm cyc}})^2\sigma_{\rm T},\nonumber\\
  &&\sigma_{\rm \bot} \approx \sigma_{\rm T}, \label{eqn03}
\end{eqnarray}
where $\sigma_{\rm T}$ is the Thomson cross-section. Thus low energy photons ($E<E_{\rm cyc}$, scattered inside the column) from the thermal mound tend to diffuse along the field line due to $\sigma_{\rm \|} < \sigma_{\rm \bot}$, i.e., forming a ``pencil-like'' beam. In the direction of the column axis a portion of the pencil-like beam is easily reprocessed by the accretion flow (see the pencil-like beam `B' in Fig. \ref{fig07}). Due to the advection of the accretion flow and gravitational light deflection, the reprocessed pencil-like beam possibly produces a narrow ``anti-pencil'' beam \citep{SasakiKKC10, SasakiMKF12}, which preserves the absorption feature in the pencil-like beam. According to the discussion in \citet{SasakiKKC10}, the anti-pencil should be from pole 2, and can be observed on the back side of the accretion column. Note that the 16 keV line appears in the anti-pencil beam ($\phi \sim$ 0.9$-$1.1, especially, see panel `004-b' of Fig. \ref{fig06}), thus the 16 keV line is formed on pole 2, and the 12 keV line is on the other pole (see Fig. \ref{fig07}).


Then we can further in TPCM explain the observations of the other harmonic components.
\begin{enumerate}[1)]
  \item{At $\phi\sim$ 0.6$-$0.9, EWs of the 12 keV line and its first harmonic ($\sim$ 22 keV) both arrive at their peaks (see panels `b' and `c' in Fig. \ref{fig06}). However, at $\phi\sim$ 0.2$-$0.5 the slight weakness of the 12 keV line is in conflict with the significant absorption near 22 keV. The discrepancy is consistent with the previous studies \citep{ArayaH00, SchonherrWE07}, i.e., it is the de-excitation of the thermal electrons and photon filling near the energy of the fundamental line that weaken the strength of the fundamental line. At $\phi\sim$ 0.2$-$0.5 the count rate in the 8$-$14 keV band is $\sim$ 15 times of that in the 19$-$27 keV band, in accordance with the property of the pencil-like beam (see the above discussion following Eq. \ref{eqn03}). Thus among photons in 8$-$14 keV, the ratio of those being absorbed is very small, as compared to the situation of photons in 19$-$27 keV, which causes a weaker absorption feature near 12 keV.}
  \item{The phase-dependent EW of the 33/35 keV line becomes complicated, in that the line is either the first harmonic of the 16 keV line, or the second harmonic of the 12 keV line. At $\phi \sim$ 0.8$-$0.9, the 12 keV line set is the main contribution to the absorption near 33/35 keV, since the 16 keV line is very weak. At $\phi \sim$ 0.1$-$0.2 or 0.5$-$0.6, the situation indicates that the 16 keV and 12 keV line sets both contribute to the formation of the 33/35 keV line, and due to a higher ratio of photons being absorbed in 30$-$40 keV EW of the 33/35 keV line is larger than that of the 12 keV or 16 keV line. Additionally, at $\phi \sim$ 0.2$-$0.5, due to lack of high-energy photons (and/or high-energy electrons) in the direction undergoing the cyclotron absorption, the absorption near 33/35 keV is very weak.}
\end{enumerate}

More observations are also consistent with the model, e.g., (i) each cyclotron line is created in a region close to the continual radiation region \citep[the height of the line-formation region is $\sim$ 0.2 km, and that of the thermal mound is $\lesssim$ 0.1 km. See][]{BeckerW07,BeckerKSE12} , in that the cyclotron line can be detected in most pulse phases (Fig. \ref{fig06}). (ii) $\frac{\Delta{B}}{\Delta{z}}$ in each line-formation region should be small, otherwise the cyclotron line would be much wider than the observed, or even hardly be observed. (iii) The configuration of the two magnetic poles in Fig. \ref{fig07}, deduced from Fig. \ref{fig06}, is consistent with that in \citet{SasakiMKF12}.

However, the strong pulse phase-dependence of $E_{\rm cyc}$ is not explicit in TPCM. It is the phase-dependent height of the line-formation-region that causes the phase-dependence \citep{StaubertTK19}. In order to figure out the dependence, more details should be considered (e.g., the light-bending around the NS), which is beyond the exploration of TPCM.

Therefore, we suppose that the two line sets (their fundamental lines are different) are formed in different magnetic poles, respectively (see Fig. \ref{fig06} and \ref{fig07}). Note that the altitude ($z$, measured from the NS surface) of the line-formation region is determined by the luminosity rather than the magnetic field \citep[see Eq. (40) in][]{BeckerKSE12}, these two lines should be produced at the same height of different poles, respectively. For ObsID 002 (004), the emission from the thermal mound undergoes the cyclotron absorption at $z\sim$ 0.23 km (0.16 km). From the centroid energies of the two fundamental lines, we obtain the magnetic fields of the two poles, i.e., $\sim$ {1.4\TDP{12}} and {1.1\TDP{12}} G, respectively.

Even though centroid energies of these two fundamental lines both seem to decrease with the decaying outburst in our work (see Tbl. \ref{tab01}), no final conclusion can be drawn. In TPCM, we can make some predictions on their luminosity-dependent $E_{\rm cyc}$. As summarized by previous studies \citep[e.g.,][]{BeckerKSE12, DoroshenkoTME17, StaubertTK19}, two types of correlations between the luminosity and CRSF energy are observed. That is, in the subcritical state a positive correlation is detected, and in the supercritical state an anti-correlation is. We suppose that at different levels of luminosities the luminosity-dependent centroid energy of the 12 keV (16 keV) line should also follow these correlations, respectively.

Future studies are expected to be performed as follows. (1) If the physical nature of the ``10 keV feature'', described by a wide-{\tt gaussian} profile or a {\tt compTT} component, is well studied, some methods should be developed to distinguish the ``10 keV feature'' from the cyclotron line in the energy spectrum. e.g., the negative (positive) dependence of the luminosity with the assumed CRSF energy in the supercritical (subcritical) state may support the feature as a cyclotron absorption \citep{ReigM16}. (2) Theoretical calculations and observations should be followed, e.g., revealing the physical properties of these different line-formation regions, measuring $E_{\rm cyc}$ of the two sets of lines at different levels of the luminosity, and disentangling the contribution of the emissions from two magnetic poles to EWs of these lines. Then we might understand why two line-formation regions appear in 4U 0115+63, and predict the same situation in other accreting pulsars. Our work is helpful to understand more issues of the cyclotron line and the distribution of the magnetic field, e.g., the luminosity-dependence of the cyclotron line energy at different levels of luminosities \citep[e.g., ][]{DoroshenkoTME17, StaubertKE17, StaubertTK19, TsygankovLCS07}. Especially, for different line sets, the luminosity-dependent line energy should satisfy different functions, respectively \citep[e.g., ][]{TsygankovLCS07, BoldinTL13}.

\subsection{Conclusion}
\label{sec04.2}

In our work, we have studied two pointing observations of 4U 0115+63 in the 2015 outburst, obtained by {\it NuSTAR}. In both observations, we have detected several harmonic CRSFs ($\sim$ 12, 23 and 33/35 keV) and a peculiar one ($\sim$ 16 keV), similar to those jointly observed in the 2011 outburst by several X-ray detectors \citep[$\sim$ 15 keV,][]{IyerMD15}. It is clear that the 16 keV line is not a harmonic component of the 12 keV line. We suppose that the fitting residual around 16 keV is not a so-called ``10 keV feature'', of which the physical nature is still an open issue \citep[e.g.,][]{StaubertTK19}. Then the robustness of the 16 keV line is confirmed. First, because of the high performance of {\it NuSTAR} the complicated cyclotron lines are detected with a higher S/N and less uncertainty, as compared to the previous observation in \citet{IyerMD15}. Secondly, in the fits using physical or phenomenological models, the absorption features near 12 keV and 16 keV are very significant and should be fitted preferentially. Thirdly, our simulations indicate that no simultaneous-detection probability of the 12/16 keV line with other lines is lower than {\DP{-5}}.

From the pulse phase-dependent equivalent widths of these cyclotron lines (see Fig. \ref{fig06}), we infer that the 12 keV and 16 keV lines are two different fundamental lines. In the two-poles CRSF model, the two line sets are produced at the same altitude ($\sim$ 0.2 km away from the NS surface) of different magnetic poles, respectively (see Fig. \ref{fig07}). Thus the magnetic fields of the two poles should be $\sim$ {1.1\TDP{12}} and {1.4\TDP{12}} G, respectively. It is expected that the centroid energy of the 12 keV (16 keV) line should satisfy the positive/negative correlation with the luminosity at different levels of luminosities, as summarized in previous work \citep[e.g.,][]{BeckerKSE12, DoroshenkoTME17, StaubertTK19}.

%


\begin{acknowledgements}

We are grateful to an anonymous referee, Prof. Fang-Jun Lu, P. A. Becker, G. K. Jaisawal and M. Yukita for clarifying and helpful comments. This work is supported by Project U1838201, National Key Research and Development Program of China (2016YFA0400803), the Natural Science Foundation of China under grant Nos. 11733009, Y71131005C, 11673023, 11333004 and 11773015 supported by NSFC and CAS. This research has made use of the {\it NuSTAR} Data Analysis Software ({\tt NuSTARDAS}) jointly developed by the ASI Science Data Center (ASDC, Italy) and the California Institute of Technology (Caltech, USA), and the software provided by the High Energy Astrophysics Science Archive Research Center (HEASARC).

\end{acknowledgements}

\bibliographystyle{apj}


\begin{figure*}
\begin{center}
  \begin{overpic}[width=0.35\textwidth, angle=-90]{f04.ps}
  \end{overpic}
  \caption{Energy spectra and spectral residuals if adding a wide {\tt Gaussian} around 10 keV in the fitting model.}
  \label{fig04}
\end{center}
\end{figure*}

\begin{figure*}
\begin{center}
  \includegraphics[width=0.45\textwidth, angle=-0]{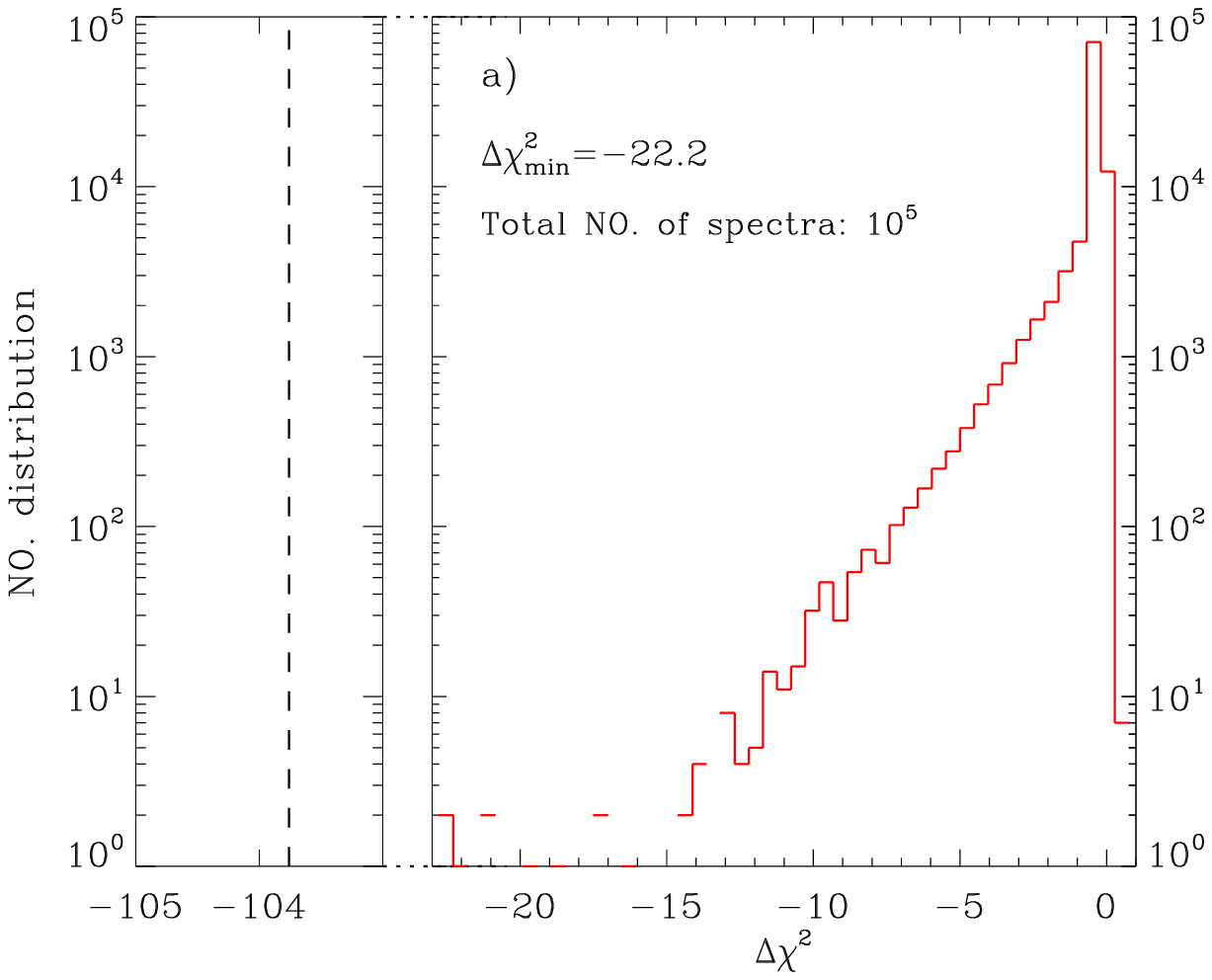}
  \includegraphics[width=0.45\textwidth, angle=-0]{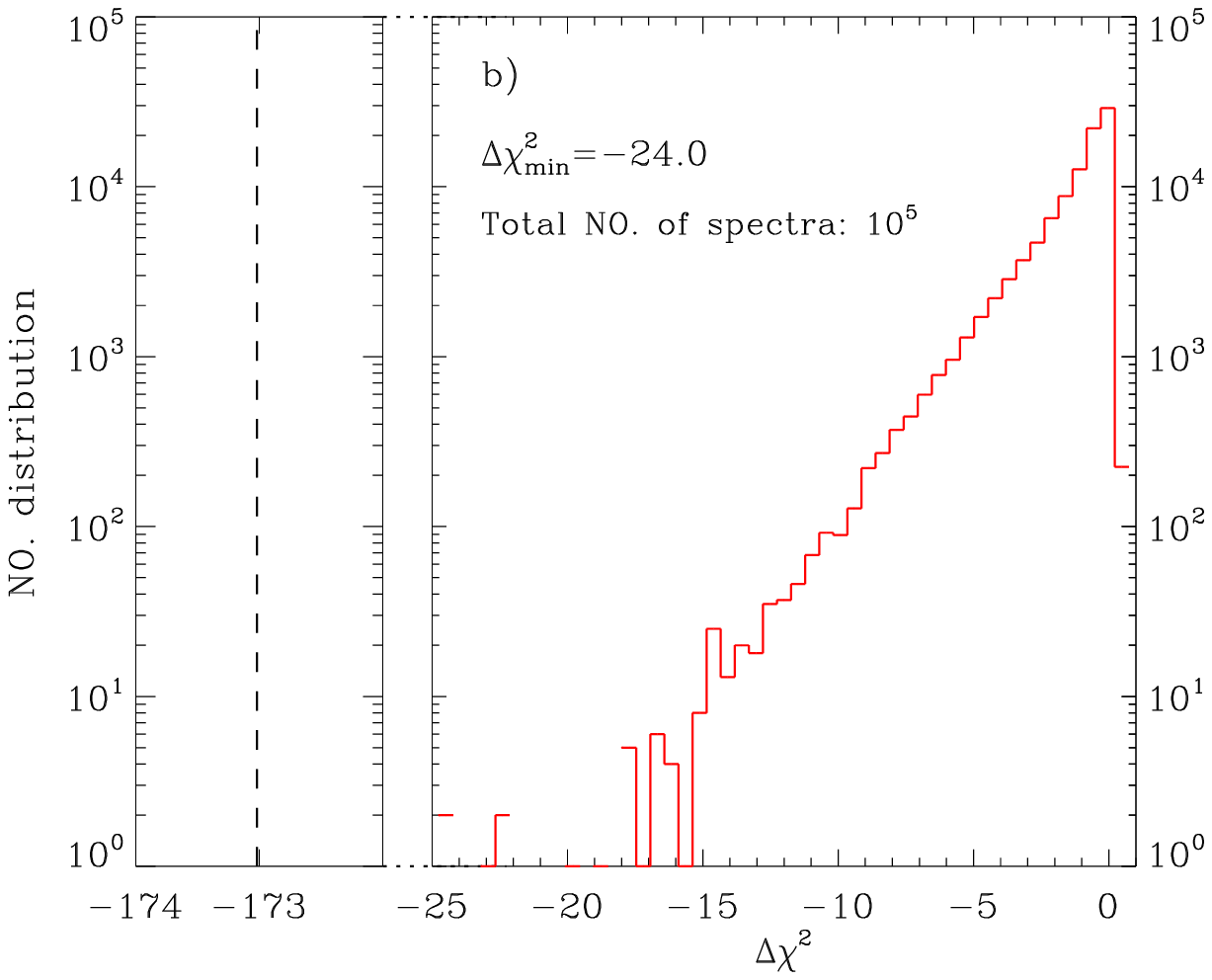}
  \caption{Results from the significance simulations. The histograms show $\Delta\chi^2$ by adding a 16 keV (left panel) or 12 keV (right panel) line to {\DP{5}} simulated spectra. The observed $\Delta\chi^2$ are marked by the dashed lines. A model consisting of {\tt compTT+gauss+bb} and 4 {\tt mgabs} components is used to simulate the spectra (see Sec. \ref{sec03.2} for more details).}
  \label{fig05}
\end{center}
\end{figure*}

\begin{figure*}
\begin{center}
  \includegraphics[scale=0.8, angle=0]{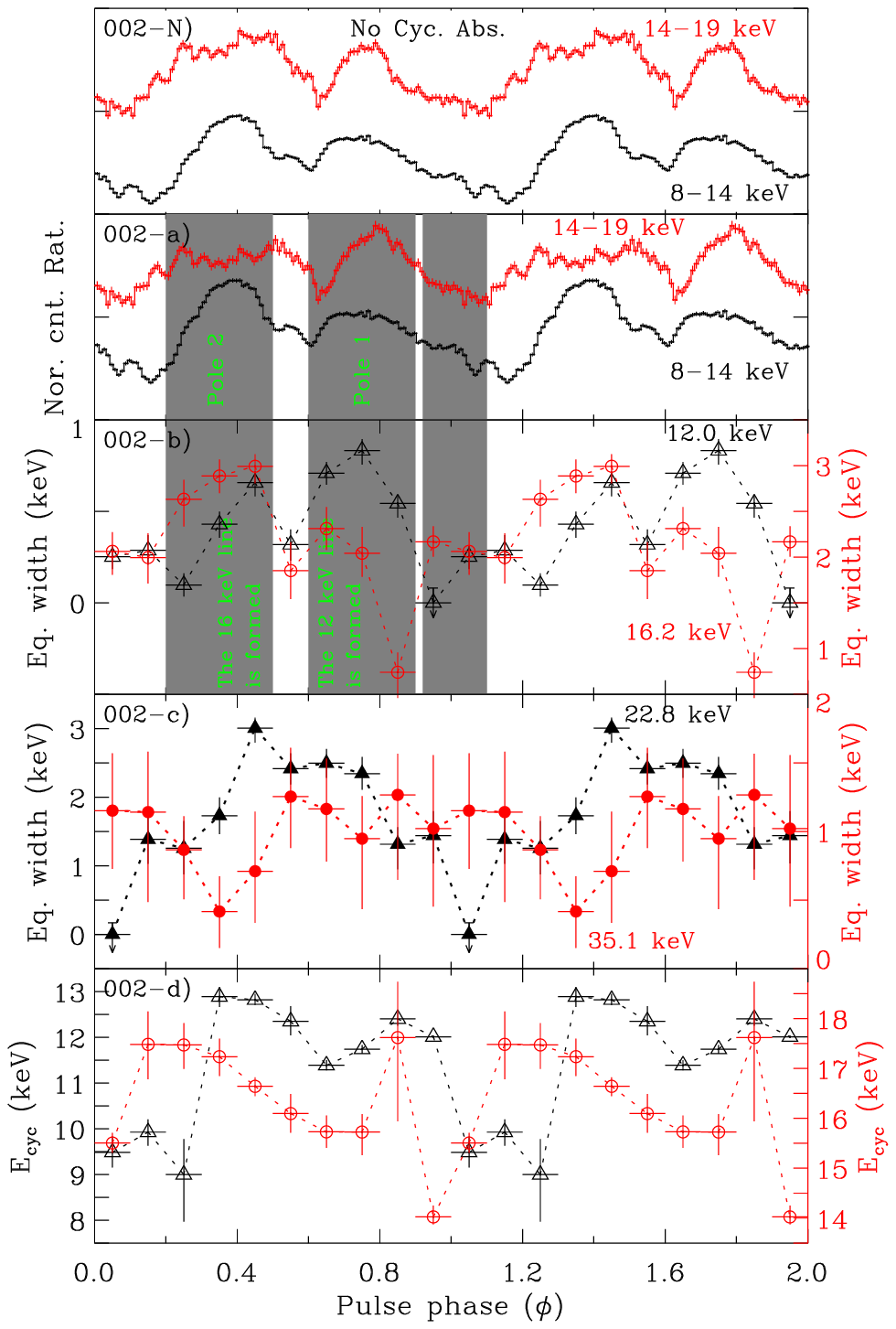}
  \includegraphics[scale=0.8, angle=0]{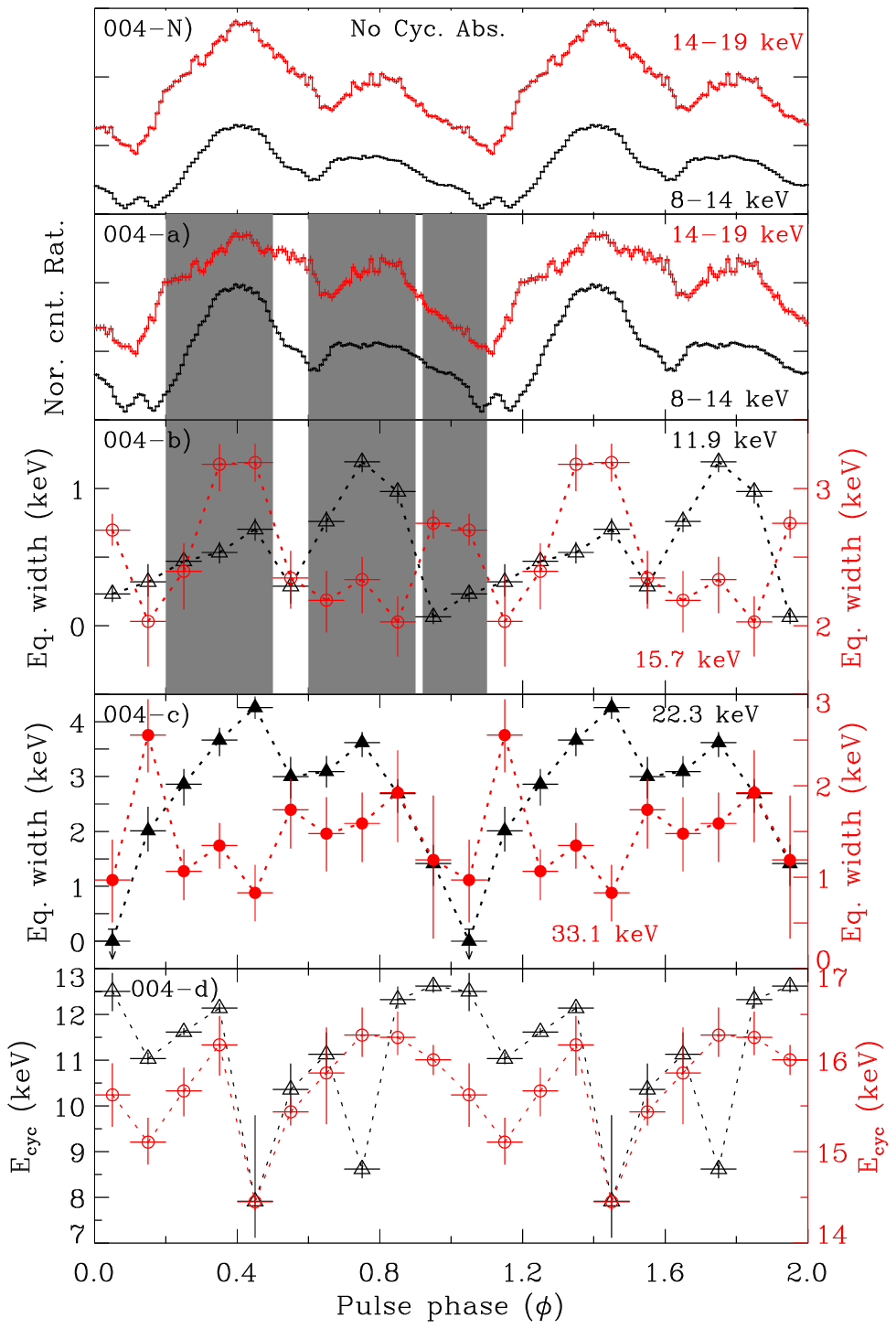}
  \caption{Pulse profiles and equivalent widths of the CRSFs for ObsIDs 90102016002 (left panels) and 90102016004 (right panels). The black open triangles, red open circles, black filled triangles and red filled circles indicate the lines of 12, 16, 22 and 33/35 keV, respectively. In some cases, we fix the centroid energy of the CRSF unless at least one of its upper and lower limits can be determined (see the point with a downward arrow). Different magnetic poles are marked in the observed pulse profiles. For comparison, we also calculate the pulse profiles without the cyclotron absorption in the top panels (panel `N').}
  \label{fig06}
\end{center}
\end{figure*}

\vspace{-2.0cm}
\begin{figure*}
\begin{center}
  \includegraphics[scale=0.8, angle=0]{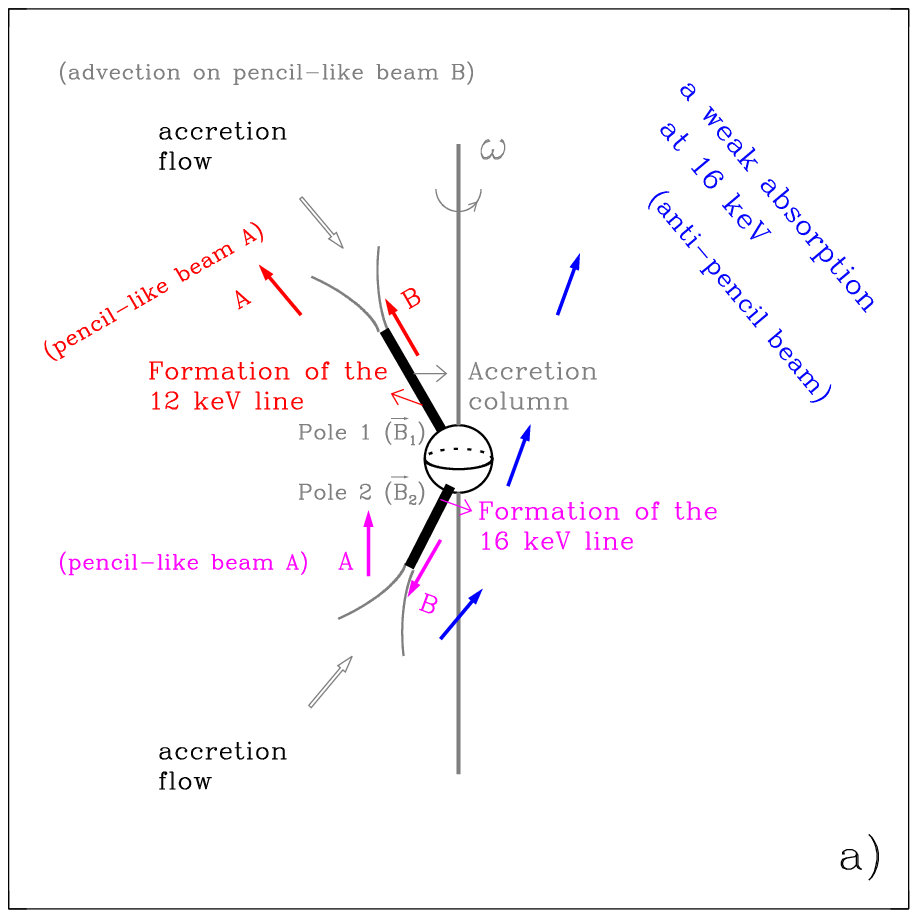}
  \includegraphics[scale=0.8, angle=0]{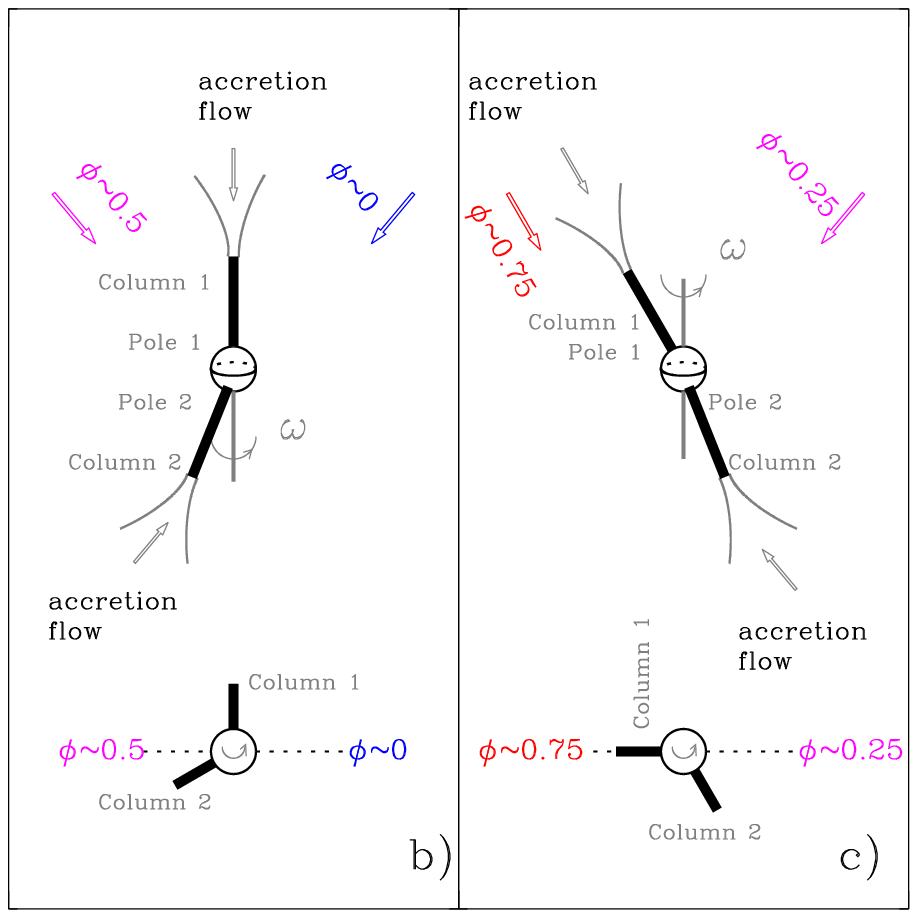}
  \vspace{0.2cm}
  \caption{Panel `a': schematic of the two line-formation regions on the two magnetic poles, respectively. Here the 12 keV (16 keV) line is formed on pole 1 (pole 2) with the field of $B_{\rm 1}$ ($B_{\rm 2}$). It is important to note that a part of the pencil-like beam (marked in `B') produces a narrow anti-pencil beam, while the rest part acts as a hollow pencil-like beam (marked in `A'). Panel `b' and `c': schematic of the phase-dependent cyclotron absorption. $\phi \sim$ 0.2$-$0.5, 0.6$-$0.9 and 0.9$-$1.1 are three key pulse phases, at which we detect two different sets of cyclotron lines, as shown in Fig. \ref{fig06}. That is, the 16 keV (12 keV) line detected at $\phi \sim$ 0.2$-$0.5 may be in the pencil-like beam  of column 2 (column 1); a portion of the 16 keV line at $\phi \sim$ 0.9$-$1.1 may be in the anti-pencil beam of column 2.}
  \label{fig07}
\end{center}
\end{figure*}

\linespread{1.4}
\begin{deluxetable}{lcc|lcc}
\tabletypesize{\scriptsize}
\tablewidth{0pt}
\tablecolumns{6}
\tablecaption{\label{tab01}Spectral parameters of the best-fitting models.}
\startdata
\tableline
\tableline
Parameters-A\tablenotemark{a} &   ObsID-002\tablenotemark{b}  &  ObsID-004               &  Parameters-B            &        ObsID-002         &  ObsID-004               \\
\tableline
$kT_{\rm 0}$ (keV)      &  2.43$^{+0.05}_{-0.03}$  &  2.66$^{+0.05}_{-0.04}$  &$kT_{\rm 0}$ (keV)      &  1.97$^{+0.03}_{-0.04}$  &  2.39$^{+0.04}_{-0.03}$  \\
$kT_{\rm e}$ (keV)      &  9.34$^{+0.28}_{-0.30}$  &  11.75$^{+0.40}_{-0.38}$ &$kT_{\rm e}$ (keV)      &  8.21$^{+0.28}_{-0.21}$  &  10.53$^{+0.40}_{-0.47}$ \\
$\tau_{\rm comp}$       &  1.69$^{+0.09}_{-0.09}$  &  0.97$^{+0.07}_{-0.07}$  &$\tau_{\rm comp}$       &  2.54$^{+0.12}_{-0.11}$  &  1.25$^{+0.12}_{-0.11}$  \\
$N_{\rm comp}$          &  0.17$^{+0.01}_{-0.01}$  &  0.09$^{+0.00}_{-0.00}$  &$N_{\rm comp}$          &  0.17$^{+0.00}_{-0.01}$  &  0.12$^{+0.01}_{-0.00}$  \\
$E_{\rm Fe}$ (keV)      &  6.51$^{+0.04}_{-0.04}$  &  6.52$^{+0.03}_{-0.03}$  &$E_{\rm Fe}$ (keV)      &  6.52$^{+0.04}_{-0.03}$  &  6.53$^{+0.03}_{-0.03}$  \\
$\sigma_{\rm Fe}$ (keV) &  0.32$^{+0.05}_{-0.04}$  &  0.26$^{+0.04}_{-0.04}$  &$\sigma_{\rm Fe}$ (keV) &  0.34$^{+0.05}_{-0.04}$  &  0.29$^{+0.04}_{-0.04}$  \\
$kT_{\rm bb}$ (keV)     &  0.67$^{+0.02}_{-0.02}$  &  0.69$^{+0.02}_{-0.02}$  &$kT_{\rm bb}$ (keV)     &  0.61$^{+0.02}_{-0.02}$  &  0.62$^{+0.02}_{-0.02}$  \\
$R_{\rm bb}$ (km)\tablenotemark{c}       &  20.63$^{+1.14}_{-1.00}$ &  15.93$^{+0.86}_{-0.72}$ &$R_{\rm bb}$ (km)       &  23.62$^{+1.89}_{-1.46}$ &  19.82$^{+1.41}_{-1.23}$ \\
$E_{\rm 1}$ (keV)       &  12.01$^{+0.17}_{-0.13}$ &  11.85$^{+0.20}_{-0.15}$ &   $E_{\rm 1}$ (keV)    &  11.79$^{+0.17}_{-0.13}$ &  11.73$^{+0.16}_{-0.14}$ \\
$\sigma_{\rm 1}$ (keV)  &  1.48$^{+0.18}_{-0.16}$  &  1.62$^{+0.18}_{-0.17}$  &$\sigma_{\rm 1}$ (keV)  &  1.50$^{+0.21}_{-0.17}$  &  4.53$^{+0.22}_{-0.17}$  \\
$\tau_{\rm 1}$          &  0.08$^{+0.02}_{-0.01}$  &  0.10$^{+0.03}_{-0.02}$  &    $\tau_{\rm 1}$      &  0.07$^{+0.02}_{-0.02}$  &  0.67$^{+0.01}_{-0.02}$  \\
$E_{\rm 2}$ (keV)       &  16.20$^{+0.40}_{-0.33}$ &  15.67$^{+0.28}_{-0.21}$ &$E_{\rm G}$ (keV)       &  8.71$^{+0.34}_{-0.37}$  &  10.06$^{+0.09}_{-0.11}$ \\
$\sigma_{\rm 2}$ (keV)  &  3.17$^{+0.43}_{-0.30}$  &  2.89$^{+0.27}_{-0.23}$  &$\sigma_{\rm G}$ (keV)  &  2.95$^{+0.19}_{-0.16}$  &  2.54$^{+0.10}_{-0.08}$  \\
$\tau_{\rm 2}$          &  0.22$^{+0.02}_{-0.01}$  &  0.30$^{+0.02}_{-0.02}$  &   $N_{\rm G}$          &  0.11$^{+0.02}_{-0.01}$  &  0.43$^{+0.02}_{-0.02}$  \\
$E_{\rm 3}$ (keV)       &  22.81$^{+0.40}_{-0.33}$ &  22.31$^{+0.32}_{-0.30}$ &$E_{\rm 2}$ (keV)       &  23.10$^{+0.38}_{-0.24}$ &  22.58$^{+0.23}_{-0.22}$ \\
$\sigma_{\rm 3}$ (keV)  &  2.80$^{+0.42}_{-0.38}$  &  3.31$^{+0.28}_{-0.33}$  &$\sigma_{\rm 2}$ (keV)  &  1.78$^{+0.50}_{-0.29}$  &  2.56$^{+0.40}_{-0.30}$  \\
$\tau_{\rm 3}$          &  0.19$^{+0.02}_{-0.02}$  &  0.25$^{+0.02}_{-0.02}$  &$\tau_{\rm 2}$          &  0.09$^{+0.01}_{-0.01}$  &  0.18$^{+0.02}_{-0.02}$  \\
$E_{\rm 4}$ (keV)       &  35.08$^{+0.30}_{-0.40}$ &  33.14$^{+0.77}_{-0.91}$ &$E_{\rm 3}$ (keV)       &  35.16$^{+0.40}_{-0.42}$ &  33.59$^{+0.90}_{-0.96}$ \\
$\sigma_{\rm 4}$ (keV)  &  1.87$^{+0.63}_{-0.55}$  &  2.36$^{+1.20}_{-0.75}$  &$\sigma_{\rm 3}$ (keV)  &  2.05$^{+0.45}_{-0.44}$  &  1.67$^{+1.35}_{-0.81}$  \\
$\tau_{\rm 4}$          &  0.16$^{+0.03}_{-0.03}$  &  0.08$^{+0.02}_{-0.03}$  &$\tau_{\rm 3}$          &  0.19$^{+0.02}_{-0.04}$  &  0.06$^{+0.03}_{-0.03}$  \\
$E_{\rm 5}$ (keV)       &  48.58$^{+2.30}_{-1.44}$ &             --           &$E_{\rm 4}$ (keV)       &  48.90$^{+1.98}_{-1.20}$ &             --           \\
$\sigma_{\rm 5}$ (keV)  &  4.06$^{+2.76}_{-1.68}$  &             --           &$\sigma_{\rm 4}$ (keV)  &  6.18$^{+3.22}_{-1.13}$  &             --           \\
$\tau_{\rm 5}$          &  0.18$^{+0.06}_{-0.06}$  &             --           &$\tau_{\rm 4}$          &  0.25$^{+0.06}_{-0.04}$  &             --           \\
$\chi^2/{\rm dof}$      &        1828.0/1731       &        1955.9/1751       &$\chi^2/{\rm dof}$      &        1824.9/1731       &        1952.2/1751       \\
Null hyp. prob.         &       5.2{\TDP{-2}}      &       4.1{\TDP{-4}}      &Null hyp. prob.         &       5.7{\TDP{-2}}      &       5.0{\TDP{-4}}      \\
\tableline
\enddata
\tablecomments{The {\tt compTT} model is used to fit the continuum. The table contains four key parameters of model {\tt compTT} (the temperature of input photons $kT_{\rm 0}$ and thermal electrons $kT_{\rm e}$, photon depth $\tau_{\rm comp}$ and normalization $N_{\rm comp}$), two of {\tt bb} (the blackbody temperature $kT_{\rm bb}$ and emission radius $R_{\rm bb}$, assuming a source distance of 7 kpc), two of Fe line (the energy $E_{\rm Fe}$ and width $\sigma_{\rm Fe}$), three of wide-{\tt gaussian} (the energy $E_{\rm G}$, width $\sigma_{\rm G}$ and normalization $N_{\rm G}$), and three of each CRSF (the line energy $E_{\rm i}$, width $\sigma_{\rm i}$ and depth $\tau_{\rm i}$ of the i-th cyclotron line, where i=1, 2, 3, 4, 5). Here errors are calculated at a level of 90\% confidence in all cases.}
\tablenotetext{a}{Parameters-A (-B) indicates the results without (with) a wide-{\tt gauss} model.}
\tablenotetext{b}{ObsID-002 (-004) stands for ObsID 90102016002 (90102016004).}
\tablenotetext{c}{$R_{\rm bb}$ is overestimated, and should be corrected using Compton scattering of the blackbody radiation \citep[$\sim$ 7 km, see discussions in][]{IyerMD15}.}
\end{deluxetable}

\end{document}